\documentclass{IEEEtran}
\usepackage{graphicx,
	psfrag,
	epsfig,
	epstopdf,
	amsthm,
	amssymb,
	url,
	subcaption,
	algorithm,
	algorithmic,
	balance,
	enumerate,
	color,
    url,
	setspace,
	tikz,
	pgfplots
}
\usepackage{amsmath}
\usepackage[nospace,noadjust]{cite}
\usepackage{pbox}
\usepackage{geometry}
\usepackage{multirow}
\usepackage{adjustbox}
\usepackage{array}
\usepackage{booktabs}
\usepackage{float}
\usepackage{tabularx}

\newcommand{\cref}[1]{Constraint~\ref{#1}}

\newcommand{\ignore}[1]{}
\newgeometry{left=1.2cm,right=1.2cm,bottom=1.5cm,top=1.5cm}

\begin{document}
\title{A Blockchain Framework for Secure Task Sharing in Multi-access Edge Computing}	

		\author{
			\IEEEauthorblockN{Angelo Vera Rivera \IEEEauthorrefmark{1}, Ahmed Refaey \IEEEauthorrefmark{2}\IEEEauthorrefmark{3}, and Ekram Hossain\IEEEauthorrefmark{1}
			\thanks{Corresponding author: Ekram Hossain (email: ekram.hossain@manitoba.ca). The work was supported in part by a Discovery Grant from the Natural Sciences and Engineering Research Council of Casnada (NSERC). }}
		
						\IEEEauthorblockA{\IEEEauthorrefmark{1} 
				University of Manitoba, Manitoba, Canada}\\
			\IEEEauthorblockA{\IEEEauthorrefmark{2} Manhattan College, Riverdale, New York, USA}\\
			\IEEEauthorblockA{\IEEEauthorrefmark{3}Western University, Toronto, Ontario, Canada}}

\maketitle

\begin{abstract}
In the context of Multi-access Edge Computing (MEC), the task sharing mechanism among edge servers is an activity of vital importance for speeding up the computing process and thereby improve user experience. The distributed resources in the form of edge servers are expected to collaborate with each other in order to boost overall performance of a MEC system. However, there are many challenges to adopt global collaboration among the edge computing server entities among which the following two are significant: ensuring trust among the servers and developing a unified scheme to enable real-time collaboration and task sharing. In this article, a blockchain framework is proposed to provide a trusted collaboration mechanism between edge servers in a MEC environment. In particular, a permissioned blockchain scheme is investigated to support a trusted design that also provides incentives for collaboration. Finally, Caliper tool and Hyperledger Fabric benchmarks are used to conduct an experimental evaluation of the proposed blockchain scheme embedded in a MEC framework.
\end{abstract}



\section*{Introduction}
Network operators and service providers are experiencing huge technical challenges to keep up with market demands that come along with the growth of new communication technologies~\cite{1}. Two examples of those challenges are the handling of increasing amount of data and the necessity of low-latency and ultra-reliable network access schemes. These have pushed operators to make huge investments in the development of new strategies to tackle these challenges.  Multi-access Edge Computing (MEC) has become a very popular framework that offers a solution to these challenges \cite{4}. According to an IDC Expenditure Report, the projected cumulative investment in communication technologies such as MEC will be about 1.7 trillion US dollars until 2020 with a peak of more than 370 US billion dollar in the year 2022. This trend is expected to continue in the years afterward \cite{15}.

MEC is a framework that allows relocation of computing and storage resources away from the cloud layer to the Radio Access Network (RAN) which is closer to the end user (i.e. at the \textit{edge} level). In fact, this change in the architecture increases the ability of an operator to adapt better to changes in the network environment, improve Quality of Service (QoS), and make the network more efficient \cite{3}. Some of the major MEC services and applications are shown in Fig.~1. Computation tasks related to many of these services and applications should be done in very short timescales and the results need to feed back to remote devices \cite{3}. A MEC framework enables the  concept of offloading computational tasks from mobile terminals with limited processing and storage to edge servers and also to data centers/clouds with enough computing and storage capabilities. The task offloading process in a MEC system can be performed in six stages: i) server discovery, ii) task segmentation, iii) migration decision, iv) task upload, v) remote task execution, and vi) return of results \cite{4}.

\begin{figure}[h]
\centering
\includegraphics[width=0.5\textwidth]{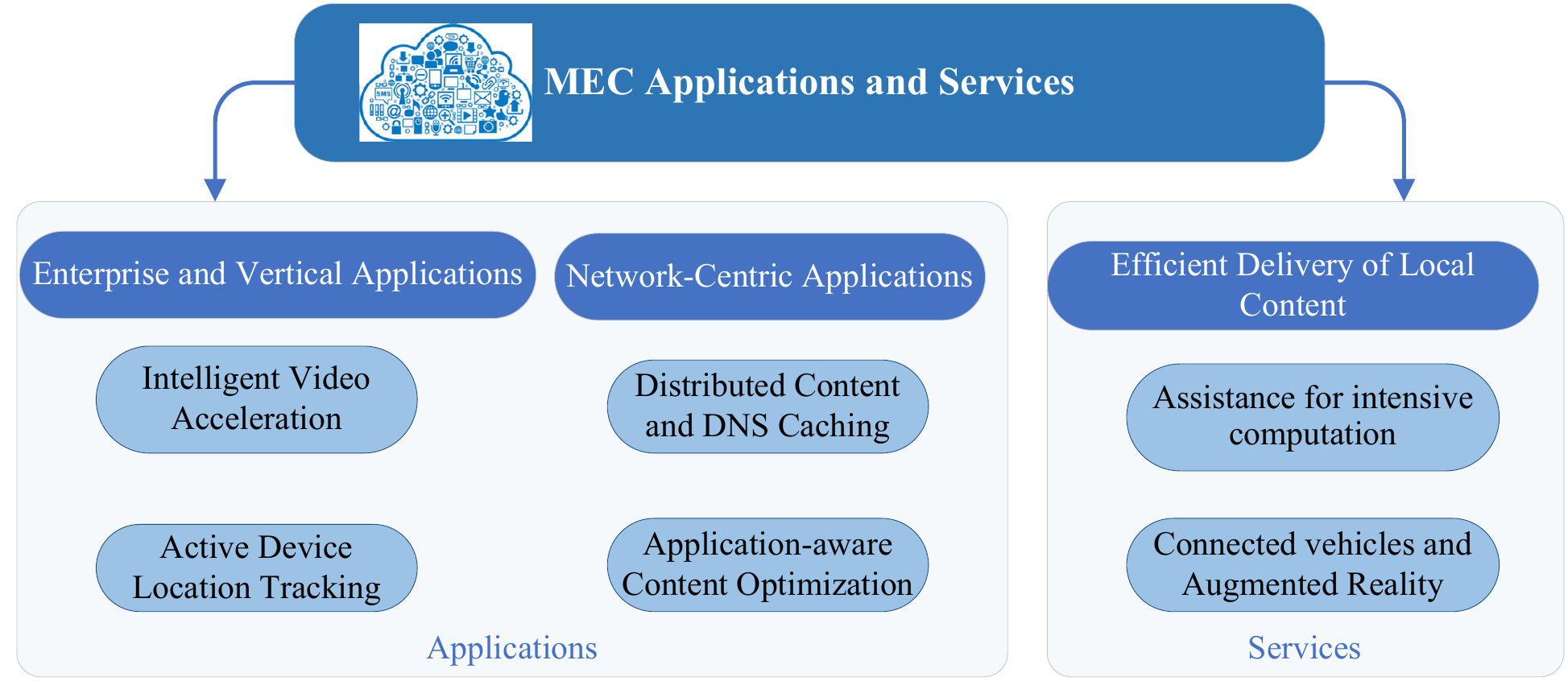}
\caption{MEC services and applications.}
\label{fig:sdparchitecture}
\end{figure}



Although the offloading approach is proved to be a worthy asset in a MEC framework, the above mentioned stages reveal several attack points that can compromise the framework during a real threat \cite {16}. In particular, the existing MEC frameworks do not address the possibility of sharing/offloading tasks among servers at the edge level. In the case that an edge server is occupied, it remains dependent on the cloud to move or share pending tasks to another edge server so they can be taken care of. This results in a deterioration of the QoS and has widespread effects on the network efficiency. In short, there are security problems during the offloading stages and there is also a lacking of task sharing exploration method at the edge level. It is important to note that most of the work on MEC in the existing literature have focused on the offloading problem from the perspective of the end devices \cite {17,9, 10}. 



In recent years, blockchain technologies, with very popular applications specially in the financial industry and crypto-currencies, have drawn a lot of attention in the research community. However, this technology has the potential to be extended to many areas beyond the financial sector and that includes communications  as well as MEC systems \cite{5}. In a very condensed way, blockchain is a decentralized ledger maintained by a group of independent users who do not trust each other. Users can make transactions with any other user in the network without supervision of a central authority. Every valid transaction is registered in the ledger and all the users gets to keep a copy of the ledger \cite{14}. This new idea of interaction between entities without control of a central authority is what makes blockchain special and interesting for MEC. However, with the rapid development of blockchain platforms, it is crucial to choose a suitable blockchain scheme for a MEC application in terms of performance, resilience, and privacy.\\ 

In this article, a task sharing mechanism based on Hyperledger Fabric blockchain is proposed for the MEC framework. In this mechanism, Hyperledger Fabric is used to add a security blanket to the task sharing/offloading processes among servers at the edge level. In particular, the use of a the task sharing/offloading mechanism at the edge level through an efficient and secure method will eliminate the need of sending the task to the cloud level in order to be executed properly. This will improve the QoS provided by the MEC framework and make the network operation more efficient. The contribution of this article can be summarized as follows:

\begin{itemize}
  \item Introduction of a task offloading/sharing mechanism among servers at the edge level,
  \item Investigation of the Hyperledger Fabric blockchain as a possible solution to account for security and privacy requirements in a tasks sharing context,
  \item Evaluation of the Hyperledger Fabric Blockchain using Caliper Benchmarking Tool to test the performance and identify potential bottlenecks.
\end{itemize}


\section*{Related Work}
The existing MEC models and blockchain-based MEC-related research work will be summarized in the following subsections.

\subsection*{Existing Models for MEC}

In traditional cloud-based centralized communications and computing schemes, network applications that reside on a cloud are allocated in clusters. A cluster is essentially a set of aggregated servers placed at centralized locations such as data centers. In that context, all the user's requests that point to a particular application are sent to the cluster it resides on in order to generate a response. However, such an implementation does not necessarily provide the best experience to end users who may be located far away from the centralized cluster. The centralized cluster approach can cause a significant decline in the end user experience due to increased delay~\cite{10}. Also, customers in the current cloud-service environments must deploy their applications to a single cloud and utilize the elasticity of the cloud to determine additional resources and spin those up accordingly within the cloud environment. If the cloud provider experience technical difficulties, the end user has no option to re-route requests to another cloud having the performance dramatically impacted. As a result, the concepts of fog Computing and edge Computing have emerged~\cite {10, 11}.


 
In a very general sense, fog Computing consists in moving intelligence power, data processing activities and data storage away from a centralized point (i.e. data center) to a local area network where the end users are located. A fog computing framework/standard, which defines the architecture for computation, storage and network services between end devices and a cloud, can describe how MEC works. In this context, MEC is a specific network architecture in which applications, services and data are pushed away from centralized points to the radio access network (RAN). This change defiantly reduces the network congestion and allow applications to perform better since the related processing tasks run closer to the end user. MEC technology is designed to be implemented at or near the cellular base stations in three scenarios as follows:
\begin{itemize}
  \item MEC at the LTE macro base station site.
  \item MEC at the multi-technology cell aggregation site.
  \item MEC at the radio network controller site.
\end{itemize}

\subsection*{Blockchain-Based MEC}
The combination of MEC and blockchain networks is emerging rapidly and has been integrated into many application scenarios such as cellular networks and device-to-device (D2D) networks. In particular, devices such as mobile units, personal computers, pads (in a Wide-Area Network [WAN]) and vehicles (in a Vehicular Ad-hoc Network [VANET]) can be equipped with blockchain clients \cite{veh}. It is important to note that in this applications the core idea of blockchain, a robust and distributed tool for secure communication, remains intact. Following this trend, the existing work on the computation offloading problem in MEC has achieved some positive results. Most of these results, however, focus either on offloading the mining tasks from the edge devices or implementation of novel incentive and reputation mechanisms. Also, a few recent studies have dealt with the comparison between multi-hop versus single-hop computation offloading problem. However, {\em all these studies focus on the offloading problem from the devices' perspective and does not address the task sharing/offloading problem among the edge servers}.

For example, in \cite{14}, a comparison among several permissioned and unpermissioned blockchains is provided. In fact, permissioned blockchains, such as a Hyperledger Fabric, can be seen as traditional blockchains with an extra security system \cite{6}. Furthermore, a permissioned blockchain allows specific actions to be performed by only certain identifiable participants of the network through the implementation of an access control layer and peer-to-peer communication. For the above mentioned reasons, Hyperledger Fabric differs from conventional blockchains. It is also a perfect fit to address the security problem in a MEC framework in the context of a task sharing mechanism between servers at the edge level.

\section*{Blockchain Basics}
A blockchain network is a distributed database or ledger maintained by a group of independent users that do not trust each other. Users in a blockchain network can make transactions among each others without supervision of a central authority through the use of cryptography. Every valid transaction is registered in the ledger and all the users get to keep a copy of it. The mechanism by which a transaction is declared valid and added to the ledger is called Consensus. Valid transactions are added to the Ledger in the form of blocks that are connected chronologically to form a blockchain. Normally, the construction of blocks and their addition to the ledger involves a consensus algorithm execution of which in some cases is a computationally difficult problem. A blockchain needs to enforce three fundamental characteristics:
\begin{itemize}
\item The ledger needs to be immutable. 
\item The ledger needs to be public and transparent.
\item The users in a blockchain are anonymous.
\end{itemize}
Even though there exist many classifications for blockchain networks, the one pertaining to Permissioned and Unpermissioned blockchains is of particular interest in the context of this work. The following subsection lays out the details of this class and sets the ground as to why this distinction might be important for MEC applications. 

\subsection*{Permissioned and Unpermissioned Blockchain}
When it comes to the access to blockchain ecosystems, these networks can be divided into permissioned and unpermissioned. Permissioned blockchains are closed environments where every participant needs to reveal full or part of its identity in order to receive permission to be part of the network. The governance aspect of permissioned blockchains is managed entirely by a group of designated nodes that makes the important decissions in the environment. Due to this governance characteristic, permissioned blockchains are often referred to as consortium. Unpermissioned blockchains, on the other hand, are completely public and any participant with a computer can join the network and be part of the decentralized community without revealing its identity. The governance aspect is highly decentralized and all the important decisions are made in a collective fashion. Unpermissioned blockchains are often referred to as public.

In terms of consensus mechanisms to make decisions inside the network, permissioned blockchains use consolidated power methods such as Proof of Stake (PoS) or Practical Byzantine Fault Tolerant (pBFT) algorithms that provide faster decision times with less computation power and lower energy consumption. In contrast, unpermissioned blockchains use public and more democratic consensus mechanisms such as Proof of Work where more nodes have a say in the decision. However, this comes at the stake of lower decision times, the need of more computation power, and higher energy consumption.

In terms of transparency, transactions in permissioned blockchains are not completely visible to the public unless some level of clearance is provided to a participant by the consortium. This hurts a bit the idea of transparency, but on the other hand, increases privacy and makes the concept more practical for some applications. When it comes to anonymity, permissioned blockchains do not enforce the idea too much since a node need to reveal full or part of its identity in order to receive permission to join the network.

In unpermissioned blockchains, transparency is strictly enforced, hence, all transactions are available and readable by every participant in the network. This characteristic is essential, because in a public unpermissioned blockchain, no central authority is implemented then transparency enforcement is a mechanism of establishing trust in the network. In terms of anonymity, technically unpermissioned blockchains guarantee complete anonymity even though in reality identity can be relevant or nor depending on the application.

As mentioned before, consensus in permissioned blockchains is reached by a few designated nodes. In addition, no mining process is involved thus making consensus mechanism in these environments computationally inexpensive, more scalable in size, and more flexible. On the contrary, unpermissioned blockchainns reach consensus in a more democratic but inefficient fashion posing serious constraints in terms of size scalability and flexibility. Table I summarizes some relevant aspects of both permissioned and unpermissioned blockchains.

\subsection*{Hyperledger Fabric}
Hyperledger is a general use consortium permissioned blockchain infrastructure that provides private transactions, confidential smart contracts and flexible pluggable consensus algorithms. Hyperledger relies on the vision that there will be a world of many blockchains in the future and Hyperledger can provide the platform where different blockchain applications could exist. The unique features of Hyperledger allow this platform to have good performance, scalability, privacy and confidentiality features. Nowadays, Hyperledger is becoming increasingly popular in many industries such as supply chain and finance but the platform has the appeal to attract a wide range of different applications and industries including communications and mobile networking and computing, as  explored in this article.

\begin{table*}[]
	\caption{\label{Table 1} Comparison between permissioned and unpermissioned blockchains}
	\centering
\begin{tabularx}{1.0\textwidth} { 
  | >{\raggedright\arraybackslash}X 
  | >{\centering\arraybackslash}X 
  | >{\centering\arraybackslash}X | }
 \hline
   & Permissioned & Unpermissioned \\
 \hline
Performance &  Consensus mechanism: PoS \& pBFT (consolidated power in a consortium) & Consensus mechanism: PoW (public power) \\
& Computation Consumption: Low & Computation Consumption: High \\ 
& Energy Consumption: Low & Energy Consumption: High \\
& Transaction Latency: Low & Transaction Latency: High \\
& Transaction Throughput: High & Transaction Throughput: Low \\
& Scalability: High & Scalability: Low (few hundred nodes) \\
& Summary: Overall efficient in terms of performance & Summary: Overall not efficient in terms of performance \\
\hline
Resilience  & Access: closed blockchain & Access: open blockchain \\
& Only nodes with credentials are allowed & Anyone can participate \\
& Security: enhanced with additional levels of access control& Security: vulnerable due to no access control \\ 
\hline
Privacy & Identity: may be known by others & Identity: completely private \\
& Role: defined by the consortium nodes & Role: no defined role \\
& Data: may be kept private and not completely accessible & Data: public and distributed among participants \\
& Terms (smart contracts): may be private & Terms (smart contracts): public\\
\hline
\end{tabularx}
\end{table*}
\section*{Blockchain-Based MEC Framework}

In the context of mobile networks, the current architectural trend tends to favor a vast number of low cost commodity network nodes. Each network node typically has limited capabilities in terms of computing power, processor speed, memory, and network bandwidth. Consequently, edge computing networks are designed to incorporate Commercial Off-The-Shelf (COTS) computers and then offload tasks to idle or underutilized systems with the ability to support peak loads.

However, in real scenarios, these computing resources are still underutilized in large percentage compared to their total service life. Subsequently, there is a developing interest for mechanisms that provide effective administration of these resources so that the promise and the vision of the MEC architecture can be fully exploited. In that direction, the benefits of blockchain technology in connection with edge computing can be summarized in terms of three relevant functions that blockchains provide:

\begin{itemize}
  \item {\em Notarization}: the tamper-proof nature of blockchain allows confirmation of authenticity and integrity of declarations by participants. 
  
 \item {\em Ownership}: the double-spending resilience provided by blockchain guarantees the consistency of the ownership history. This ensures that there is a unique owner at any given point in time.
  \item {\em Provenance and Chain of Custody}: cryptographic identifiers for items can hold item's origin and history of custody in a way that allows subdivision and combination of items.
\end{itemize}

\subsection*{Proposed Network Model}
The proposed system combines the open standards and management technology of Hyperledger Fabric blockchain with edge computing to provide a powerful and lightweight extensible technology foundation for securely managing edge computing servers. The proposed model is built on top of Hyperledger Fabric for two main reasons: 
\begin{itemize}
  \item \textbf{Firstly}, Hyperledger is different from other blockchains because it allows point-to-point communication. Precisely, every individual has a set of ``states" on their copy of the ledger. The ledgers between nodes are actually different in that only those parties that are privy to information or transactions can see the shared ``state". Also, Consensus is two fold: i) it uses contracts which validate input and outputs to transactions as well as required signatures, and ii) it uses a notary service to avoid double spends as the notary is included on all transactions involving a certain ``state".
  \item  \textbf{Secondly}, Hyperledger is also a permissioned type of blockchain that does not implement a mining concept to achieve consensus. In particular, every new transaction is broadcast to all peers and consensus is achieved via the Practical Byzantine Fault Tolerance (PBFT) algorithm. Then, each of the peers executes the transaction and generates a block.
\end{itemize}

The main goal of the proposed system is to offer a service-oriented architecture (SOA) approach for edge computing servers in order to securely share tasks on network nodes. To achieve this goal, the edge servers are instrumented in such a way that Hyperledger Fabric is responsible for their effective monitoring and management. In particular, the MEC system integrates the Hyperledger Fabric protocols designed for peer-to-peer secure communication with remote management tools and mechanisms that can run on the cloud level. 

On the cloud level, the system enables remote management of the network's resources through a dynamic monitoring and metering of the requested services such as sharing tasks. In that direction, the system is adapted to dynamically discover network resources, peer association, binding of communication between peers, and provision of services for network edge nodes. On the edge server level, on the other hand, the system provides dynamically distributed services to the peer nodes in the network. Each of these nodes is enabled with components that expedite the use of Hyperledger Fabric blockchain. In particular,  the MEC framework will support distributed computing with dynamic task scheduling and unified management of resources while maintaining distributed security capabilities with a special focus on integrity and confidentiality of data and performed tasks.

As a result, the proposed system will provide a smart, model-driven, distributed, and heterogeneous communication architecture between the edge and cloud levels. In particular, it creates intelligent capabilities for autonomous collaboration among the edge nodes as follows:
\begin{itemize}
  \item {\em Autonomy of edge nodes}: edge servers can perform autonomous connection, discovery, learning, and execution of tasks through the integration of Hyperledger Fabric capabilities.
  \item {\em Collaboration among edge nodes}: Edge servers can share tasks among themselves and also between them and the cloud. This provides elastic networking, computing, and storage capabilities.
\end{itemize}

\subsection*{Sharing Framework}
In the proposed system, the relationship among different edge server nodes changes from master-slave to equal partnership through the peer-to-peer communication in the Hyperledger Fabric. Also, the most appropriate node to process a task is dynamically determined based on its willingness to tackle the task and its availability. This selection will involve a series of optimizations running on the cloud level. Consequently, the cloud level will have a system view of all the candidate nodes that volunteer to accept the task. In addition to that, the task itself will not be released to all of the candidate nodes or the cloud. Instead, it will be shared only between the node requesting the service and the candidate node selected to tackle the task through the peer-to-peer communication in the Hyperledger Fabric platform.

\begin{figure*}[h]
\centering
\includegraphics[width=1\textwidth]{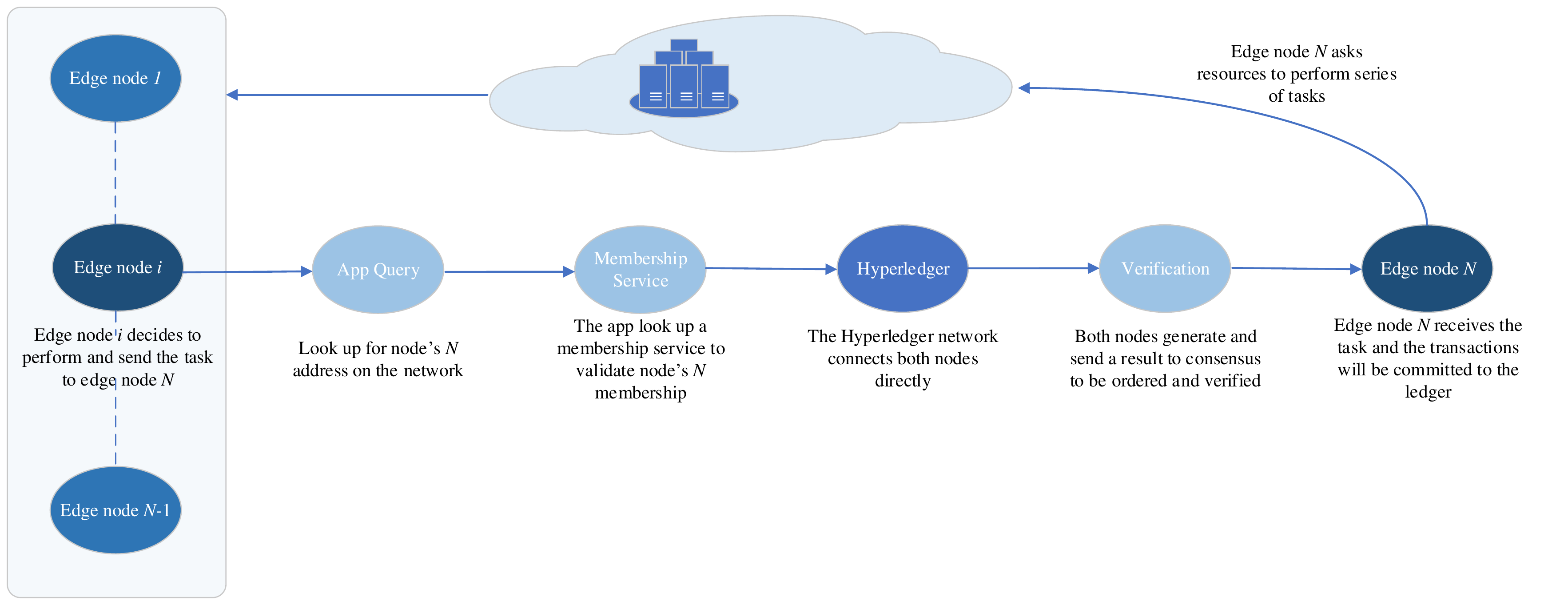}
\caption{Service request and secure task sharing work flow.}
\label{fig:sdparchitecture}
\end{figure*}
\section*{Hyperledger Fabric Architecture for MEC}
The proposed system architecture (in Figure \ref{fig:sdparchitecture}), it is assumed that a Hyperledger Fabric network runs with  $M$ number of nodes. Node $N$ requests additional resources to perform a specific task, and the cloud level is assumed to have a monitoring tool in addition to a participating node within the network. Whereas the cloud level has a system overview, it is also able to shortlist some edge servers (i.e. Node 1 -- Node $N$-1), all of which are capable of performing this task based on the information available through the monitoring tools. The request of a service and the posterior task sharing will follow the sequence below:
\begin{enumerate}
 \item Node $N$ requests a service (sharing task(s)) from the cloud level.
  \item The cloud level runs a monitor tool, therefore, it has a shortlist of the best resources which are available to tackle this task.
   \item The cloud level will contact the shortlisted resources to check if any of them is willing to execute the task based on their local situation.
   \item The node which is willing to execute the task, Node $i$, will start a contract with Node $N$.
   \item The sharing process will follow in order within the Hyperledger Fabric framework.
\end{enumerate}
It is important to note that the nodes directly affiliated to the transaction (i.e. sharing task) are connected, while the ledger that receives updates regarding this transaction includes only Node $i$ and Node $N$. This dynamic ensures confidentiality and privacy during the process. As shown in Figure \ref{fig:sdparchitecture}, the transaction flow begins with the edge node placing all requested task sharing into a block as a transaction. Then, this node will sign the batch and send it to the cloud level to validate. Once in the cloud level, the application in charge of validation will use its transaction processor to ensure the integrity of the batch and then commit it.


\subsection*{Work Flow}
The type of smart contracts used in the proposed system is Installed Smart Contacts (ISC). The ISC's will include the logic running on the blockchain and the validators in the network before it is launched. Also, the ISC's are responsible for processing all transaction requests including the validation process with reference to the logic. As controlled by the implemented policy in the system, any edge node can activate a smart contract by submitting an instantiating transaction to the network. When it is approved, the smart contract enters into an active state so that it can receive transactions from users. Any validated transaction is appended to the shared ledger. The contact will be individual through standalone instances.

\begin{figure}[h]
\centering
\includegraphics[width=.5\textwidth]{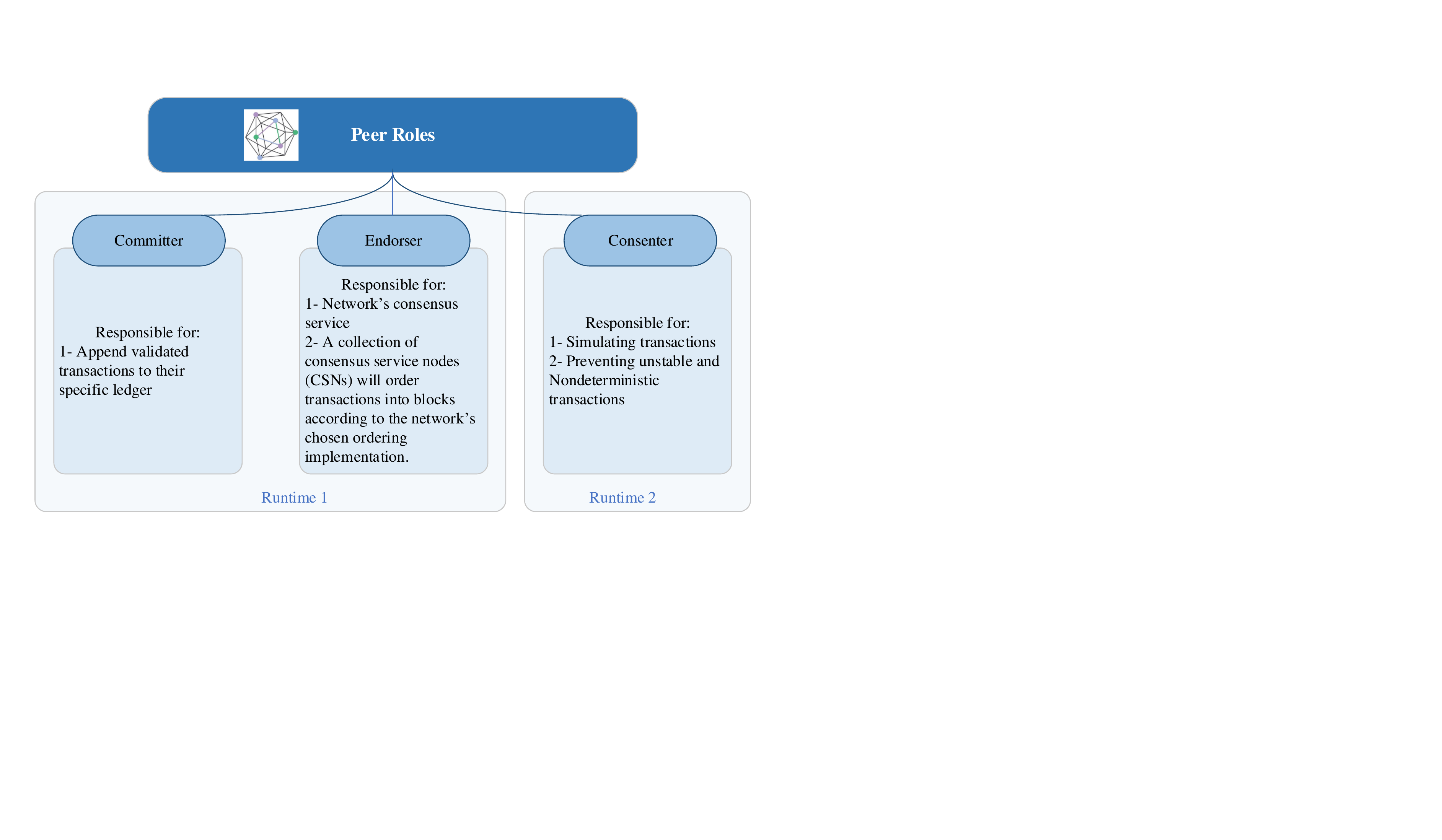}
\caption{Peer roles in Hyperledger Fabric.}
\label{Hyperledger Peers}
\end{figure}

As shown in Figure \ref{Hyperledger Peers}, peers are separated into two different run times and three distinct roles: Endorsers, Committers, and Consenters. When Endorsers and Committers appear, they are executed on the same run-time and Consenters are run on a completely separate run-time. This allows the Hyperledger Fabric modular architecture to enable the consensus mechanism to be a plug-and-play feature which in return allows a high degree of customization of the network according to the needs of the business. The three types of peers in Hyperledger Fabric and their run-times are described as follows:

\begin{enumerate} 
\item Run-time 1: This run-time includes the Endorser and Committer Peers. It is not necessary for the committer to have chain-code installed, however, they maintain full ledger of records. This means that Committers are incapable of running smart contract functions. In contrast, all Endorser peers must have chain-code installed and prepares the transaction proposal based on smart contract results.
\item Run-time 2: This run-time includes Consenter which simulate transactions in an isolated chain-code container.
\end{enumerate}

In fact, there are three types of peers: Endorsing, Committing, and Ordering peers. The Ordering peers have a role of receiving endorsed transactions and packaging them into blocks as per the configuration file. Then, they send these blocks to all other peers to validate those transactions and update their ledgers. Furthermore, the Ordering peers keep track of all transactions in their ledgers (valid and/or invalid). However, endorsing and committing peers' ledgers only contain valid transactions.

\subsection*{Processing}
The Hyperledger Fabric assigns network roles based on the node type. Also, the transaction execution, ordering, and commitment are separated to provide parallelism to the network. In fact, ordering transactions after executing them enables the peer node to process several transactions simultaneously.  This parallel execution increases the processing efficiency of each peer and accelerates the delivery of transactions to the ordering service. Furthermore, the division of labor unburdens ordering nodes from the demands of transaction execution and ledger maintenance enables parallel processing. At the same time, the peer nodes are freed from ordering (i.e. the consensus) workloads. The bifurcation of roles limits the required processing for authentication. In particular, the peer nodes do not have to trust all ordering nodes and vice versa. Therefore, processes on one can run independently of verification by the other.

\section*{Performance Evaluation}
In the context of Blockchain platforms, a performance evaluation is the process of measuring performance metrics of a system under test (SUT) \cite{14}. In a very general sense, typical measures of interest are often related to response times (latency) and throughput. Benchmarking, on the other hand, is the process of defining standard metrics and scenarios in order to make fair comparisons among different versions of a single system or across systems. In this section, the Caliper Blockchain Evaluation Tool is presented as a platform for performance evaluation of different blockchain solutions. It supports different SUT's such as Hyperledger Fabric, Hyperledger Burrow, Hyperledger Composer, Ethereum, FISCO BCOS, Hyperledger Iroha and Hyperledger Sawtooth. Some of the  performance metrics used by Caliper for the benchmarking process are Read Latency, Read Throughput, Transaction Latency and Transaction Throughput \cite{18}. A brief explanation of these metric is presented next.

Read Latency (RL) is defined as ${\rm RL} = \mbox{ReplyReceivedTime - SubmitTime}$, i.e. the time difference between when a read request is submitted and the reply is received. This metric is expressed in seconds [s]. 


Read Throughput (RT), on the other hand, is defined as ${\rm RT} = \frac{\mbox{Total Read Transactions}}{\mbox{Total Time}}$. The ratio of total read operations per unit of time. This value is expressed in Reads Per Second [rps]. The Transaction Latency (TL) is defined as ${\rm TL} = \mbox{Transaction Confirmation Time - Submit Time}$, i.e. the time difference between the confirmation time of a transaction and the submission time. It is important to note that in a blockchain environment the confirmation of a transaction happens when the transaction is added to a block and it is available in the network after the consensus mechanism has been applied. When a transaction is broadly available to the network it is said to be committed. TL is expressed in seconds [s]. Finally, Transaction Throughput (TT) is defined as ${\rm TT} = \frac{\mbox{Total Committed Transactions}}{\mbox{Total Time}}$, the ratio of total committed transactions to the blockchain network per unit of time. This rate is expressed in transactions per second [TPS]. In the figures below, an evaluation of Hyperledger Fabric versions 1.4.0 and 1.4.1 using Caliper Benchmark Tool is presented \cite{18}. The bemchmark used for this test is Fabric Marbles with StateDB set to GoLevelDB. The total number of transactions pushed into the Hyperledger is 10000 and the Input Transaction Rate increases from 500 TPS to 6000 TPS. The parameters being assessed are transaction latency, CPU usage and memory usage.\\

\begin{figure}[h]
\centering
\includegraphics[width=0.5\textwidth]{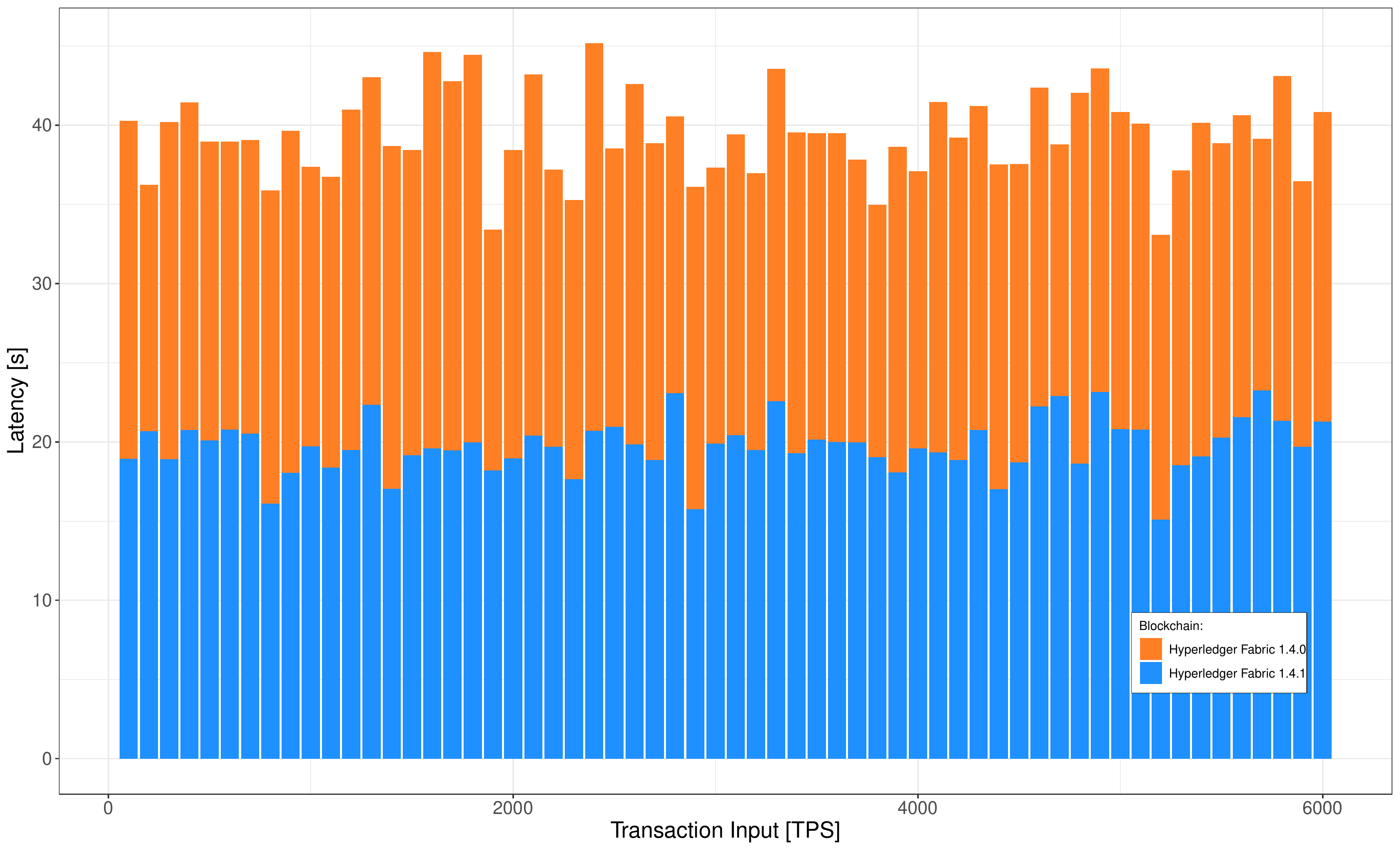}
\caption{Variation of transaction latency versus input transaction rate for Hyperledger Fabric versions 1.4.0 and 1.4.1.}
\label{evaluation 2}
\end{figure}

\textbf{Observation 1}: Figure \ref{evaluation 2} shows the behavior of Transaction Latency measured in seconds [s] with respect to Input Transaction Rate which is given in transactions per second [TPS]. For every TPS point, the average of the latency of transactions is measured. It can be observed that, Transaction Latency for Hyperledger 1.4.0 values varies from 15 seconds to 25 seconds in a quite disturbed form, whereas the Transaction Latency values for Hyperledger 1.4.1 remains more stable in this regard. Although the dissimilar variation form, the average of the latency of transactions is approximately 20 seconds for both systems under test. 

\begin{figure}[h]
\centering
\includegraphics[width=0.5\textwidth]{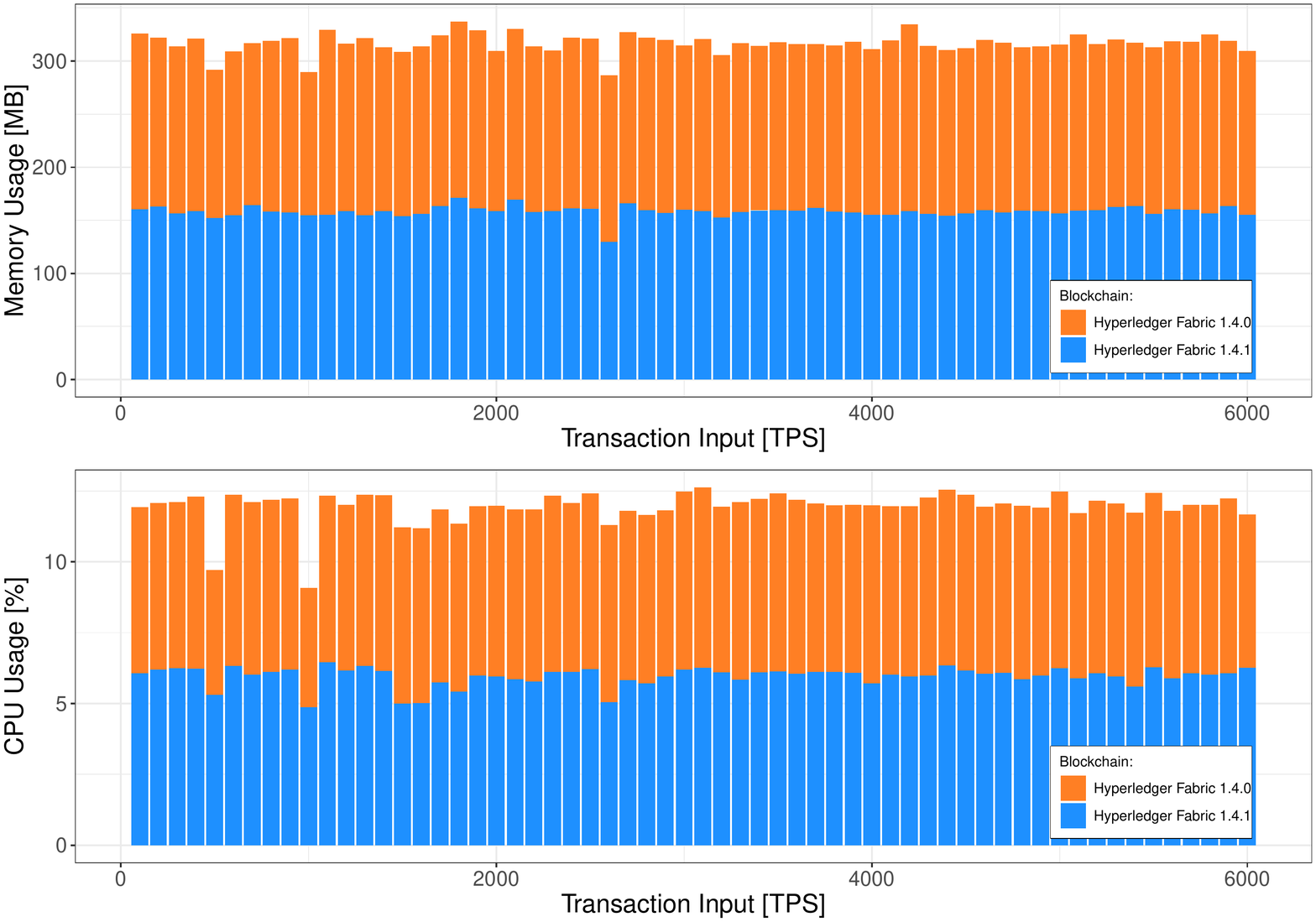}
\caption{Variations of CPU usage \& memory usage with input transaction rate for Hyperledger Fabric versions 1.4.0 and 1.4.1.}
\label{evaluation 4}
\end{figure}

\textbf{Observation 2}: Figure \ref{evaluation 4} top shows the evaluation of the Memory Usage of the system in megabytes units [MB]versus Input Transaction Rate given in transactions per second [TPS]. For every TPS point, the average of the use of memory is measured. It can be observed that for both Hyperledger Farbric 1.4.0 and 1.4.1, the Memory Usage behaves similarly across the evaluation interval with a very stable form and an average of approximately 158 MB for both systems under test.\\

\textbf{Observation 3}: Figure \ref{evaluation 4} bottom shows the evaluation of the CPU Usage measured as a percentage [\%] versus Input Transaction Rate given in transactions per second [TPS]. For every TPS point, the average of the use of CPU is measured. Similar to Observation 2, it can be examined that for both Hyperledger Farbric 1.4.0 and 1.4.1, the CPU Usage behaves almost identically with an average of 5.9\% CPU usage for both systems under test. 

\section*{Conclusion}
A blockchain-based task sharing framework is proposed for MEC servers in order to enable trustworthy and efficient collaboration. This framework will enable the end device the offloading and efficient sharing of underutilized resources. The evaluation results show that the Hyperledger Fabric version 1.4.1 provides the best CPU and memory usage, and also the transaction latency. This evaluation shows that the overhead of the proposed system is negligible in comparison to the gained benefits. The benefits are gained by enabling task offloading/sharing among the edge servers while maintaining a proven unbreakable security and privacy standards.

There are many opportunities to further enhance the proposed framework. The possible improvements are as follows:
\begin{itemize}
  \item {\em Security and task privacy}: Indeed, a blockchain framework contains a strong authentication mechanism which will be maintained at the edge server level. However, the privacy protection of the end-user is still an open problem. Furthermore, monitoring and auditing the sharing process is required to mitigate the privacy leaking to prevent attacks from malicious users and intruders \cite{x}.
  \item {\em Mobility support for end devices}: Break-off may happen when an end-user leaves the current edge server and moves to another one. The smooth migration of the task ownership among different entities is also a challenge. This can be addressed through a global view from the cloud level. 
  \item {\em Global deployment and collaboration}: For a successful  deployment, it is important to introduce a reference architecture to enable collaboration across different MEC service providers. A blockchain-based agreement is feasible to facilitate such collaboration \cite{14}.
\end{itemize}

\small
\bibliographystyle{IEEEtran}

\bibliography{bib}

\end{document}